\documentclass[sn-mathphys,Numbered]{sn-jnl}

\usepackage{graphicx}%
\usepackage{multirow}%
\usepackage{amsmath,amssymb,amsfonts}%
\usepackage{amsthm}%
\usepackage{mathrsfs}%
\usepackage[title]{appendix}%
\usepackage{xcolor}%
\usepackage{textcomp}%
\usepackage{manyfoot}%
\usepackage{booktabs}%
\usepackage{algorithm}%
\usepackage{algorithmicx}%
\usepackage{algpseudocode}%
\usepackage{listings}%

\theoremstyle{thmstyleone}%
\theoremstyle{thmstyletwo}%

\theoremstyle{thmstylethree}%

\begin{document}

\title[Article Title]{Hybrid Quantum Neural Network Structures for Image Multi-classification}


\author{\fnm{Mingrui} \sur{Shi}}\email{shiyue519416@163.com}

\author{\fnm{Haozhen} \sur{Situ}}\email{situhaozhen@gmail.com}

\author*{\fnm{Cai} \sur{Zhang}}\email{zhangcai@scau.edu.cn}

\affil{\orgdiv{College of Mathematics and Informatics}, \orgname{South China Agricultural University}, \orgaddress{\street{No. 483 Wushan Road, Tianhe District}, \city{Guangzhou}, \postcode{510642}, \state{Guangdong}, \country{China}}}


\abstract{Image classification is a fundamental computer vision problem, and neural networks offer efficient solutions. With advancing quantum technology, quantum neural networks have gained attention. However, they work only for low-dimensional data and demand dimensionality reduction and quantum encoding. Two recent image classification methods have emerged: one employs PCA dimensionality reduction and angle encoding, the other integrates QNNs into CNNs to boost performance. Despite numerous algorithms, comparing PCA reduction with angle encoding against the latter remains unclear. This study explores these algorithms' performance in multi-class image classification and proposes an optimized hybrid quantum neural network suitable for the current environment. Investigating PCA-based quantum algorithms unveils a barren plateau issue for QNNs as categories increase, unsuitable for multi-class in the hybrid setup. Simultaneously, the combined CNN-QNN model partly overcomes QNN's multi-class training challenges but lags in accuracy to superior traditional CNN models. Additionally, this work explores transfer learning in the hybrid quantum neural network model. In conclusion, quantum neural networks show promise but require further research and optimization, facing challenges ahead.}




\maketitle
\section{Introduction}\label{sec1}

Image classification is a crucial task in the field of machine learning. With the widespread application of digital images and the generation of large-scale datasets, significant breakthroughs have been achieved in image classification using deep learning. Deep learning leverages multi-layered neural network structures to automatically extract features and classify images, enabling the learning of higher-level abstract features from raw pixel-level data, thereby enhancing the accuracy and performance of image classification\cite{lecun1998gradient, krizhevsky2012imagenet, simonyan2014very, szegedy2015going, he2016deep}. The success of deep learning can be attributed to its powerful model representation capabilities and the training on massive datasets. By stacking deep neural networks, deep learning learns multi-level feature representations, including low-level edges, textures, and higher-level semantic information such as shapes and structures. Specially designed structures such as Convolutional Neural Networks (CNNs) and pooling layers enable deep learning to effectively handle spatial locality and translation invariance in images\cite{ngiam2010tiled, zeiler2014visualizing, yu2015multi, howard2017mobilenets}.

Quantum Machine Learning (QML) is an advanced research field that combines quantum computing with machine learning, aiming to leverage the characteristics of quantum computation to enhance the processing capabilities of conventional machine learning algorithms\cite{biamonte2017quantum, havlivcek2019supervised}. In classical computers, the time complexity for handling large data vectors is typically polynomial or exponential. However, quantum computing offers significantly higher storage availability than classical computing, enabling easy storage and processing of massive data. With the rapid progress of quantum computing technology, there is an urgent need to explore how to maximize its advantages across various application domains. Implementing quantum computational features in machine learning will lead to revolutionary changes in computer vision, particularly in dealing with text, images, videos, and other areas. The applications of quantum machine learning can be categorized into two primary aspects: optimizing traditional machine learning algorithms and addressing situations where classical models are unable to perfectly represent image feature correlations. Quantum machine learning is based on quantum physics principles, integrating quantum computing with machine learning to harness quantum devices' potential for enhancing traditional machine learning algorithm performance. By utilizing quantum computing concepts to improve machine learning algorithms, we can enhance existing classification and clustering algorithms. In quantum machine learning, amplitude encoding can be employed for data preprocessing, transforming classical machine learning data into formats suitable for quantum device processing\cite{schuld2018supervised, larose2020robust}.

The Variational Quantum Algorithm (VQA) is a type of quantum algorithm used in Quantum Machine Learning (QML) to solve optimization problems\cite{peruzzo2014variational, mcclean2016theory, kandala2017hardware}. As a specific application of VQA, the Parametric Quantum Circuit (PQC), also known as the Variational Quantum Circuit, is a quantum circuit composed of unitary gates with freely adjustable parameters. With the development of Parametric Quantum Circuits, researchers have started exploring their integration with machine learning to address classical problems\cite{havlivcek2019supervised, kandala2017hardware}, such as numerical optimization, approximation, classification, and more\cite{sim2019expressibility, hubregtsen2020evaluation}. Due to the resemblance of Parametric Quantum Circuits to traditional neural networks, they can be optimized by adjusting the parameters to approximate the target function. Consequently, combining Parametric Quantum Circuits with neural networks has led to the development of Quantum Neural Network (QNN) algorithms\cite{farhi2018classification, schuld2020circuit, beer2020training}. One key difference between QNN and neural networks is that all parameter gates in QNN are reversible. Additionally, PQCs do not employ activation functions for non-linear operations; instead, they use entangling layers to perform entanglement operations on the output quantum states, achieving non-linear computations.

Building upon the foundation of Quantum Neural Networks (QNNs), Cong et al.\cite{cong2019quantum} were inspired by Convolutional Neural Networks (CNNs) and proposed the Quantum Convolutional Neural Networks (QCNN) model. This model combines and enhances multiple quantum techniques, such as multiscale entanglement and quantum error correction, to process input data with fewer network parameters. The QCNN has been demonstrated to be effectively trainable on near-term quantum devices. Furthermore, researchers have explored the integration of quantum circuits and CNNs. YaoChong Li et al.\cite{li2020quantum} introduced an innovative approach by integrating parametric quantum circuits with CNNs, constructing a recognition model with quantum convolutional layers and quantum classification layers, which achieved remarkable results in experiments. In the domain of molecular modeling, K.T. Schütt et al. \cite{schutt2018schnet} employed continuous-filter convolutional layers to model local correlations and applied it to SchNet for modeling quantum interactions within molecules, enabling joint prediction of total energy and interatomic forces. Tomohiro Mano and Tomi Ohtsuki analyzed three-dimensional wave functions using image recognition methods based on multilayer CNNs\cite{ohtsuki2020drawing}. They found that training the network at the center of the band allows obtaining a complete phase diagram. Alexey A. Melnikov et al.\cite{melnikov2019predicting} proposed a graph-based machine learning algorithm, where a designed convolutional neural network learns to recognize graphs with quantum advantages without executing quantum walks or random walk simulations. Zhang et al.\cite{zhang2019quantum} introduced a quantum-based subgraph convolutional neural network structure that captures both global topological and local connectivity structures in graphs, effectively characterizing multiscale patterns present in the data. These studies provide novel ideas and approaches for the field of quantum machine learning and demonstrate feasibility and effectiveness through experimental and numerical simulations.

Existing research has shown that in many cases, quantum algorithms outperform classical computing algorithms\cite{du2020expressive, potempa2022comparing}, but it has not been proven that quantum computing can improve performance in all areas. Therefore, a significant amount of research is currently focused on identifying fields where quantum machines' parallel computing capabilities can lead to advancements. In conclusion, quantum machine learning is a challenging yet promising field. Leveraging the power of quantum computing can significantly enhance the performance of applications using quantum machine learning, driving progress in various domains. However, quantum machine learning is still in a rapid development phase, requiring further research and practice to address technical challenges. As hardware devices continue to improve, research in the quantum machine learning field will also advance.

The main focus of this paper is to investigate the strengths and weaknesses of several commonly used Hybrid Quantum Neural Network (HQNN) algorithms\cite{arun2023quantum}. There are two main approaches studied: one using Principal Component Analysis (PCA) as the dimensionality reduction method with angle-encoded QNN models, and the other combining traditional CNN models with QNN models to form hybrid quantum neural networks. The first approach is primarily applied to image multi-classification problems, followed by a comparison with the second approach. Additionally, we propose a new PQC construction approach by combining new quantum entanglement strategies and better combinations of quantum parameter gates. Furthermore, we study whether the HQNN network model improves in performance after applying transfer learning.
\section{Related Work}\label{sec2}

\subsection{Amplitude Encoding and Angle Encoding}\label{subsec21}

In machine learning, the representation of data plays a crucial role in determining the final training effectiveness of algorithms. In classical machine learning problems, data comes in various forms, including language, images, and various information components, and is often represented as one-dimensional vectors or multidimensional arrays. Such data representations facilitate appropriate processing by algorithms and help achieve desired algorithmic outcomes. However, in quantum machine learning, the representation of data differs from the traditional approach. For quantum machine learning to handle classical data, we need to find suitable methods to encode it into a quantum system. This process is referred to as quantum data encoding or quantum data embedding. The quality of data encoding directly influences the effectiveness of quantum machine learning. Several quantum data encoding methods have been proposed, such as amplitude encoding and angle encoding\cite{cong2019quantum, schuld2021effect, abbas2021power, lloyd2020quantum, taigman2014deepface}.

For a classical dataset $X = { x^{(1)},...,x^{(i)},...,x^{(m)}}$ containing $m$ samples, where each sample is a vector with $n$ features, $x^{(i)} = {x^{(i)}_1, x^{(i)}_2,...,x^{(i)}_n}$, the goal of amplitude encoding is to encode the feature data of each sample into the amplitudes of a quantum state. We can use $N = \log_2(n)$ quantum bits (qubits) to encode each data sample as follows:

\begin{equation}
    |\psi^{(i)}\rangle = \sum\limits_{j=1}^n x^{(i)}_j|j\rangle \label{eq1}
\end{equation}

It is important to note that the amplitudes of each computational basis in this quantum state must satisfy the normalization condition, and in Formula~\ref{eq1} it is required that $\sum\limits_{j=1}^n (x^{(i)}_j)^2 = 1$. After encoding the data into the amplitudes of the quantum state using amplitude encoding, subsequent algorithms will involve computations on these quantum state amplitudes. The advantage of amplitude encoding lies in significantly reducing the number of quantum bits required for data encoding, meeting the stringent requirements of NISQ devices for the number of qubits. However, it is worth mentioning that this encoding method may have lower efficiency in the preparation process, which is the process of preparing the quantum state with the desired amplitudes.

For angle encoding, each feature of the data sample is encoded into the rotation angle of each quantum bit. For the angle encoding of $x^{(i)}$, we require $n$ quantum bits to complete the encoding:

 \begin{equation}
     |\psi^{(i)}\rangle = \bigotimes \limits_{j=1}^n R(x^{(i)}_j) \label{eq2}
 \end{equation}
 
In angle encoding, $R$ represents single-qubit rotation gates, which typically include $R_x$, $R_y$, and $R_z$ gates. Angle encoding only uses $n$ quantum bits and a constant-depth quantum circuit, making it suitable for current quantum hardware. However, for samples with a large number of features, it is not feasible to directly encode the data using angle encoding. To address this issue, a common approach is to use feature dimensionality reduction techniques, which reduce the number of features to meet the requirement of the available number of quantum bits.
\subsection{Quantum Neural Network}\label{subsec22}

As a quantum machine learning algorithm, Quantum Neural Network is a model that combines quantum computing with neural networks. Its main purpose is to leverage the advantages of quantum computing to handle and learn complex data patterns. Similar to classical neural networks, a Quantum Neural Network consists of input, intermediate, and output layers. In QNN, the input layer serves as the quantum data encoding layer, the intermediate layer represents the parametric quantum circuit layer, and the output layer corresponds to the quantum measurement layer.

A parameterized quantum circuit is a quantum computing model in which the parameters of quantum gates can be adjusted based on the problem at hand. These parameterized gates can be optimized according to the input data to perform specific quantum computing tasks. Typically, a parameterized quantum circuit consists of a series of quantum gates with adjustable parameters. These parameters can be real numbers and can be trained using classical optimization algorithms to maximize or minimize specific quantum measurement outcomes. By tuning these parameters, a parameterized quantum circuit can represent a wide range of quantum operations and quantum state transformations. Parameterized quantum circuits find broad applications in quantum machine learning and quantum optimization.

In quantum computing, commonly used measurement bases include the Z-basis, X-basis, Y-basis, etc. The quantum measurement layer in a Quantum Neural Network often selects the Z-basis measurement, which corresponds to measuring the quantum state in the Z-direction and obtaining the expectation value. The expectation value can be understood as the average of measurement outcomes obtained in multiple Z-basis measurements. Therefore, the output of the Quantum Neural Network is the expectation value obtained by measuring the quantum state in the Z-basis. The expectation value of the quantum state, as the output result, possesses some characteristics:

1. Real-valued: The expectation value of a quantum state is a real number that represents the probability distribution in a specific measurement basis.

2. Interpretability: The expectation value reflects the probability distribution of a quantum state in a specific measurement basis, which helps us understand the model's processing result on input data.

3. Comparability: The expectation values of different samples can be compared, enabling tasks such as classification, regression, etc.

\begin{figure}[h]%
\centering
\includegraphics[width=0.9\textwidth]{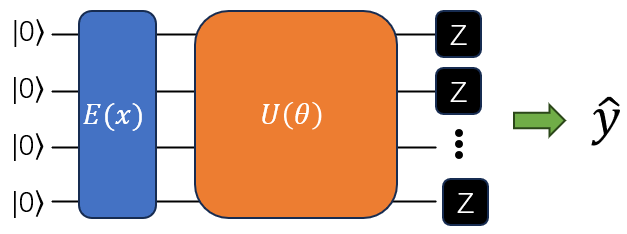}
\caption{a diagram of the quantum neural network. The quantum state is initialized as $|0\rangle$, and $E(x)$ represents the quantum data encoding layer, which is the implementation to load classical data into the quantum circuit. $U(\theta)$ denotes the parameterized quantum circuit, where $\theta$ represents the adjustable parameters in the circuit. The final layer marked with Z represents the Z-basis measurement in the quantum neural network. Finally, we obtain the output $\hat y$ of the QNN.}\label{fig1}
\end{figure}

Indeed, the output layer of a quantum neural network utilizes a quantum measurement layer, and the output result is represented by the expectation value of the quantum state. By choosing different measurement bases, one can obtain expectation values along different directions, enabling the processing and prediction of input data.

As shown in Figure ~\ref{fig1}, the role of the Quantum Data Encoding Layer $E$ is to encode classical data $x^{(i)}$ into the quantum circuit for further processing by subsequent layers. The PQC (Parameterized Quantum Circuit) layer $U$ performs complex quantum computations on the input quantum state. Finally, the output is obtained by measuring the quantum state using the Quantum Measurement Layer, and the expectation value is used as the network's output result. When we choose the Z-basis as the measurement basis, the output of the Quantum Neural Network (QNN) can be expressed as:

\begin{equation}
    \hat y^{(i)} = \langle 0 | E^{\dagger}(x^{(i)}) U^{\dagger}(\theta) Z U(\theta)E(x^{(i)}) | 0 \rangle \label{eq3}
\end{equation}

After obtaining the output vector $\hat{y}$, further processing can be performed on it.

Quantum Neural Networks are a practical application of variational quantum algorithms in quantum computing.
\subsection{Deep learning for image classification}\label{subsec23}

Deep learning has made significant progress in image classification tasks, and Convolutional Neural Networks are among the most commonly used and successful deep learning models. The design of CNNs is inspired by the working principles of the human visual system, allowing them to automatically learn and extract features from images, resulting in high accuracy in image classification.

A Convolutional Neural Network consists of multiple layers, and its key components include the convolutional layer, pooling layer, and fully connected layer. In CNNs, the convolutional layer plays the role of feature extraction. It performs convolutional operations with a set of learnable convolutional kernels on the input image to capture local features in the image. The convolutional operation slides a window over the image and performs element-wise multiplication and accumulation with the convolutional kernel to generate feature maps. By combining multiple convolutional kernels, the network can learn more abstract and high-level feature representations. Following the convolutional layer, the pooling layer acts as a downsampling method and preserves essential information. The pooling layer aggregates features from local regions (max pooling or average pooling), reducing the size of the feature maps. This helps to reduce the number of parameters, improve computational efficiency, and enhance the model's invariance to translation, scaling, and rotation transformations. The fully connected layer is situated between the convolutional layer and the output layer. In the fully connected layer, each neuron is connected to all neurons in the previous layer. By learning the weight and bias parameters, the fully connected layer transforms the feature maps extracted by the convolutional and pooling layers into the final class predictions. The fully connected layer establishes connections between high-level abstract features and specific classes, enabling classification decisions.

The success of deep learning in image classification relies heavily on large-scale annotated image datasets and powerful computational resources. By feeding image data into deep learning models and optimizing them through backpropagation algorithms, the models can automatically learn and adjust parameters to achieve optimal performance in image classification tasks. The application of deep learning in image classification extends beyond traditional object recognition and encompasses more complex tasks such as scene understanding, facial recognition, image generation\cite{everingham2010pascal, radford2015unsupervised, china1}.
\subsection{Hybrid Quantum Neural Network}\label{subsec24}

Hybrid quantum neural network is a machine learning algorithm that combines quantum neural network algorithms with classical algorithms. Its purpose is to leverage the quantum computing properties of QNN in classical algorithms or to use classical algorithms to assist QNN in accomplishing its target tasks\cite{cong2019quantum, china1, henderson2020quanvolutional, bokhan2022multiclass, trochun2021hybrid}. There are common forms of HQNN, such as using QNN as the training model and employing traditional optimization algorithms like stochastic gradient descent and Adam optimization; alternatively, QNN can be integrated as a component in combination with a traditional convolutional neural network to form a new model called CNN-QNN, which is then optimized using classical optimization algorithms\cite{pramanik2022quantum, abdel2021quantum, yang2021decentralizing}.

In the first type of HQNN model, the main component is composed of QNN\cite{li2022image}. Therefore, the primary concern is how to encode classical data into the QNN model. This encoding method is known as data encoding, and angle encoding and amplitude encoding are the main choices. Amplitude encoding requires fewer quantum bits for data with a large number of features, making it a preferred choice. On the other hand, angle encoding consumes a significant number of quantum bits, so to avoid resource consumption, dimensionality reduction techniques like linear discriminant analysis or principal component analysis are commonly used. The most crucial step in training the model is to optimize it based on its performance. The optimization process aims to find the optimal model parameters that minimize the loss function, leading to more accurate predictions on the training data and better generalization ability. Since the main component of the QNN model is the parameterized quantum circuit, we can use traditional optimization techniques to adjust its parameters and ensure the model's convergence.

In the second type of HQNN model, known as the CNN-QNN model, QNN is added as a component to the CNN network. Typically, QNN is added in the middle of the hidden layers, after the convolutional blocks, similar to a fully connected layer, and sometimes as the output layer of the entire model. For this type of HQNN model, the training process also falls into two categories. We divide the model into the CNN part and the QNN part. Without a doubt, the parameters in the QNN part will be adjusted during training optimization. As for the CNN part, the first case involves training and optimizing it together with the QNN part. In the second case, transfer learning is applied to the CNN part, where a pre-trained CNN model is transferred to the HQNN model without requiring additional training, reducing the training cost and improving efficiency to some extent.
\section{HQNN Model Design for Image Multi-Classification}\label{sec3}

\subsection{HQNN Model Design with PCA Dimensionality Reduction}\label{subsec31}

In the current mixed quantum neural network, a common approach is to incorporate QNN as a component into the neural network or directly use QNN as the target training model. Regardless of the approach, a significant challenge is how to handle the large amount of data with a limited number of quantum bits. Since image data contains numerous features, a common solution is to use PCA dimensionality reduction in combination with angle encoding. This process transforms the high-dimensional data of classical image data into low-dimensional data that can be encoded with a small number of quantum bits. Subsequently, the low-dimensional data is input to the quantum circuit using angle encoding.

\begin{figure}[h]%
\centering
\includegraphics[width=0.9\textwidth]{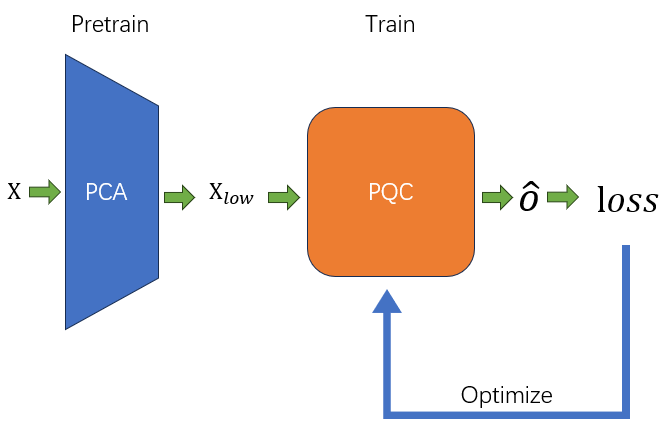}
\caption{Schematic Diagram of HQNN based on PCA Processing.}\label{fig2}
\end{figure}

As shown in Figure ~\ref{fig2}, one common construction method for the QNN model involves encoding the low-dimensional data $x_{low}$ obtained through PCA dimensionality reduction into the quantum circuit. Each value of the low-dimensional data is used as the rotation angle for the Ry gate. The subsequent part is a typical parameterized quantum circuit. The PQC starts with a set of rotation gate operations, consisting of Rx, Ry, and Rz gates, applied to each input qubit. Then, quantum entanglement operations are performed on the qubits. In the entanglement part, we require each qubit to undergo an entanglement operation. This is achieved by using controlled gates to connect different qubits. There are four common entanglement strategies: linear entanglement, cyclic entanglement, star entanglement, and fully connected entanglement\cite{sim2019expressibility}. In this design, we use the most commonly used cyclic entanglement strategy, where each qubit is entangled with its next nearest neighbor qubit. After the entanglement part, another set of rotation gate operations is typically applied to each qubit. It is important to note that the parameters of these gates are different from the previous ones, representing a new set of operations. We consider the first rotation operation, entanglement operation, and second rotation operation together as one layer, and we can repeat this layer to create a multi-layered quantum circuit. It is worth noting that the second rotation operation is not necessary in every layer; it is only needed after the final layer. Finally, we measure the expectation value of each qubit in the Z basis and use the expectation values from each qubit as our output vector $\hat o$.
\subsection{HQNN model based on CNN}\label{subsec32}

We can interpret the HQNN model based on CNN from two perspectives.

\begin{figure}[h]%
\centering
\includegraphics[width=0.9\textwidth]{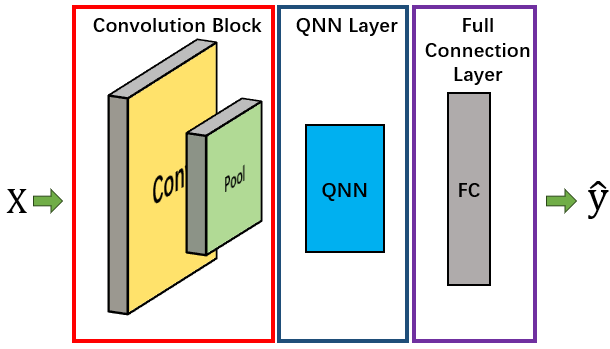}
\caption{HQNN Model based on CNN Architecture. X represents the input data. The red box represents the model's convolutional block, which includes convolutional layers and pooling layers. The subsequent blue box represents the QNN layer, and the purple box represents the fully connected layer. Finally, $\hat y$ denotes the model's output}\label{fig3}
\end{figure}

First Perspective - CNN as the Main Model Body with QNN Embedded as a Layer. Convolutional Neural Networks (CNN) are currently the most effective and practical method for processing image data. CNN has given rise to numerous models designed for various image-related tasks in computer vision, such as LeNet, VGG, ResNet for image classification, YOLO for object detection\cite{redmon2016you}, and U-net for semantic segmentation\cite{isensee2019no}. The goal of the HQNN model based on CNN is to modify the well-designed CNN model by incorporating the Quantum Neural Network (QNN) as a quantum circuit layer, referred to as the "quantum layer," and adding it to the original model, as depicted in Figure ~\ref{fig3}. There are two common approaches in current practices: one is directly adding the quantum layer to the CNN network, either between fully connected blocks or as a new output layer; the other is to replace one or more fully connected layers, utilizing the quantum layer's output either as the final output or an intermediate layer. In our illustration, we adopt the approach of adding the quantum layer after the convolutional blocks and before the output layer. This method aims to optimize and enhance the original CNN model, resulting in improvements reported in various articles.

Second Perspective - QNN as the Main Model Body with CNN as Preprocessing. In this perspective, the QNN serves as the main model body, and CNN is viewed as a preprocessing step for the QNN model. The key structure of CNN is the convolutional layer, which extracts features from input image data to obtain useful and high-level feature representations. Through end-to-end training, CNN automatically learns feature representations, eliminating the need for laborious and subjective manual feature engineering. Moreover, CNN, with convolution and pooling operations, takes advantage of local perception and parameter sharing mechanisms, effectively capturing local patterns and textures in images, which PCA often fails to do. Additionally, CNN's multi-layer convolution and pooling operations enable multi-scale representations, capturing both fine details and overall information in images, while PCA can only extract global information. The introduction of non-linear activation functions in CNN enhances the model's representational capacity, enabling it to better model complex image features and relationships. As a result, CNN provides a robust foundation of features for computer vision tasks. After feature extraction by CNN, we continue to use QNN for the classification task on the obtained feature data. Since QNN's output is constrained, we can add a final fully connected layer to control the output.

In conclusion, the HQNN model based on CNN is a novel approach that combines CNN and QNN, aiming to improve the performance of CNN or serve as a feature extraction step to fulfill QNN's requirements for low-dimensional features. The combination of these two models has shown promising results in various research studies.
\subsection{The Proposed Model}\label{subsec33}

In quantum machine learning, amplitude encoding has some advantages over angle encoding. Firstly, amplitude encoding uses the amplitudes of quantum bits to represent information, which allows it to accommodate more information. In contrast, angle encoding only utilizes the phase information of quantum bits, resulting in limited information capacity. Secondly, amplitude encoding exhibits a certain tolerance to noise and errors. Since amplitude encoding utilizes the amplitudes of quantum bits, it can retain more information even in the presence of some level of noise and errors. On the other hand, angle encoding is more susceptible to noise and errors, which may lead to information loss and misinterpretation\cite{schuld2019quantum}. Furthermore, amplitude encoding is more straightforward to implement during computation and operations. It can leverage common quantum gate operations for manipulation and measurement, whereas angle encoding requires more quantum gate operations, making it more complex to implement. Considering the large amount of data in image information, using amplitude encoding is more suitable for the current quantum device environment. It can mitigate the impact of noise to some extent and enable the storage of more information.

\begin{figure}[h]%
\centering
\includegraphics[width=0.9\textwidth]{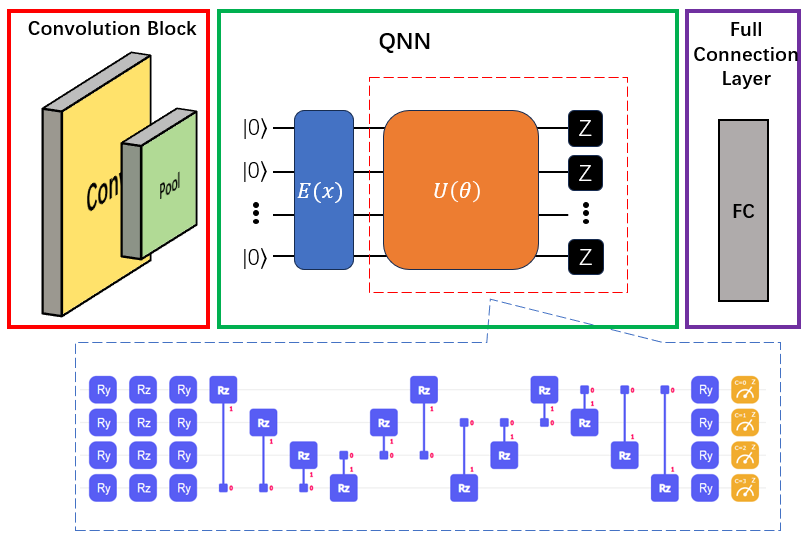}
\caption{Overall Model Diagram. In the diagram, the blue dashed box represents an example of a PQC with four qubits.}\label{fig4}
\end{figure}

In the QNN model, one of the most crucial challenges is the construction of Parameterized Quantum Circuits. Obtaining a PQC with better simulation performance and higher expressive power has been a longstanding problem in the field of QNN. To address this issue, we have incorporated the concept of a single quantum neural network simulating an approximate function of a single variable\cite{yu2022power}. This approach introduces a more expressive parameterized rotation gate U, which is illustrated in Figure ~\ref{fig4} and can be expressed concisely as follows:

 \begin{equation}
     U(\theta) = R_y(\theta_1)R_z(\theta_2)R_y(\theta_3) \label{eq4}
 \end{equation}

The parameter vector $\theta$ is represented as $\theta = (\theta_1, \theta_2, \theta_3)$ for gate U. After passing through this combination gate, the initial quantum bits exhibit a broader and more random distribution on the Bloch sphere mapping. As a result, the PQC has a higher expressive power.

Furthermore, apart from focusing on innovative ways to design parameter gates, we should also pay attention to the entanglement capability of each quantum bit\cite{sim2019expressibility}. Two common quantum entanglement strategies are full connectivity and cyclic connectivity, and full connectivity has some important advantages over cyclic connectivity. Full connectivity enables direct entanglement between all quantum bits, establishing a global entanglement relationship where each bit is directly connected to every other bit. This entanglement strategy exhibits strong entanglement capability, flexibility, and information transfer efficiency. It can construct entanglement networks of different scales, catering to various quantum computing tasks and algorithm requirements. Full connectivity also efficiently transfers and exchanges quantum information, providing faster and more efficient information transfer. Moreover, full connectivity exhibits good scalability, allowing easy expansion to larger quantum systems to meet complex computational demands.

Finally, we continue to use the conventional method of calculating the expectation value of Z-basis measurements as the output of our QNN.

Simultaneously, we embed the QNN into a simple CNN model to leverage the powerful feature extraction capabilities of CNN. After the QNN, we add a fully connected layer to obtain a more flexible model output. The overall model is depicted in Figure ~\ref{fig4}.
\section{Experiment and Result Analysis}\label{sec4}

In order to explore the performance of the model in image classification, we used two of the most commonly used datasets in the fields of machine learning and deep learning for image classification: the MNIST dataset and the FashionMNIST dataset. These two datasets are benchmark datasets in the computer vision domain. The MNIST dataset contains handwritten digit images with 10 classes, each representing a digit from 0 to 9. The images are grayscale and have a size of 28x28 pixels, with a total of 60,000 training samples and 10,000 test samples. The MNIST dataset is widely used to validate and compare the performance of various image classification algorithms, and it is considered a relatively simple dataset. On the other hand, the FashionMNIST dataset is a more challenging dataset, consisting of images of 10 categories of fashion items such as shirts, pants, and shoes. Compared to the MNIST dataset, the FashionMNIST dataset contains more complex and diverse image styles, making it more relevant to real-world applications. It also includes 60,000 training samples and 10,000 test samples, with images of size 28x28 pixels. In our experiments, we used conventional training optimization methods for all models. The training process was set to 50 epochs, and the commonly used cross-entropy loss function was used as the loss function for image classification tasks. We employed the Adam optimizer with a learning rate of 0.01. The main focus of our analysis is to compare the training process and the final test accuracy of the models to evaluate their performance.

While binary classification has achieved 100\% accuracy on many models, we are more interested in multi-class classification problems. Multi classification better reflects the complexity and diversity of the real world. Many image datasets involve multiple categories, such as object recognition, face recognition, scene classification, etc., which require more fine-grained classification of images to obtain more accurate results. Multi classification provides richer information and better meets the needs of real-world applications. For example, suppose we want to build an image recognition system that can recognize different types of animals. If we simplify the problem to binary classification, such as distinguishing between cats and dogs, then we will not be able to distinguish other animal categories, such as birds, elephants, lions, etc. However, through multi classification, we can train the model to distinguish more animal categories, giving it a broader range of applications.

\begin{table}[h]
\caption{Results of PCA Multi Classification}\label{tab1}
\begin{tabular*}{\textwidth}{@{\extracolsep\fill}lccccc}
\toprule%
& \multicolumn{4}{c}{PCA+QNN} & \multicolumn{1}{c}{New HQNN} \\\cmidrule{2-5}\cmidrule{6-6}%
Class & \multicolumn{2}{c}{2} & \multicolumn{2}{c}{8} & 10 \\\cmidrule{2-3}\cmidrule{4-5}\cmidrule{6-6}%
Dimension & 8 & 10 & 8 & 10 & None \\
\midrule
Loss  & 0.3940 & 0.4096 & 1.1819 & 1.1123 & 0.4086 \\
Accuracy & 0.8011 & 0.7954  & 0.2324 & 0.2245 & 0.8439 \\
\botrule
\end{tabular*}
\end{table}

\begin{figure}[h]%
\centering
\begin{minipage}[c]{0.5\textwidth}
\centering
\includegraphics[height=5cm,width=7cm]{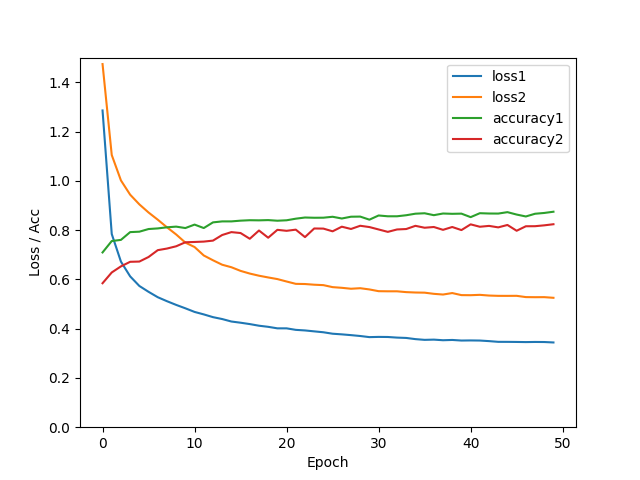}
\end{minipage}%
\begin{minipage}[c]{0.5\textwidth}
\centering
\includegraphics[height=5cm,width=7cm]{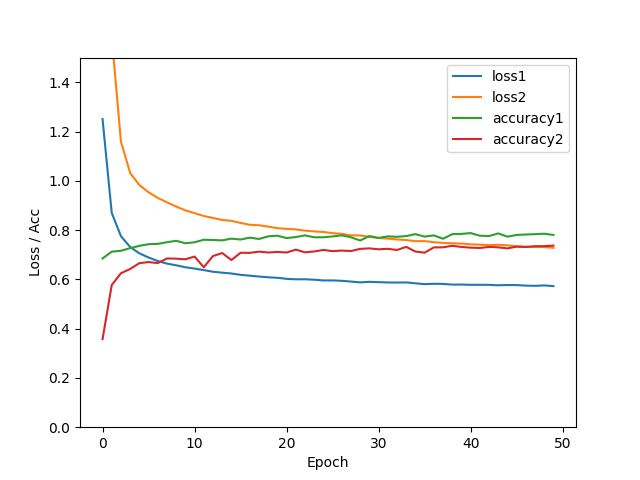}
\end{minipage}
\caption{Comparison between the Model Built with the New PQC Approach and the Traditional PQC Model. The curves "loss1" and "accuracy1" represent the loss and accuracy of the new model, while the curves "loss2" and "accuracy2" represent the loss and accuracy of the original model. The left plot shows the training results on the MNIST dataset, while the right plot shows the results on the FashionMNIST dataset.} \label{fig5}
\end{figure}

\begin{figure}[h]%
\centering
\includegraphics[width=0.9\textwidth]{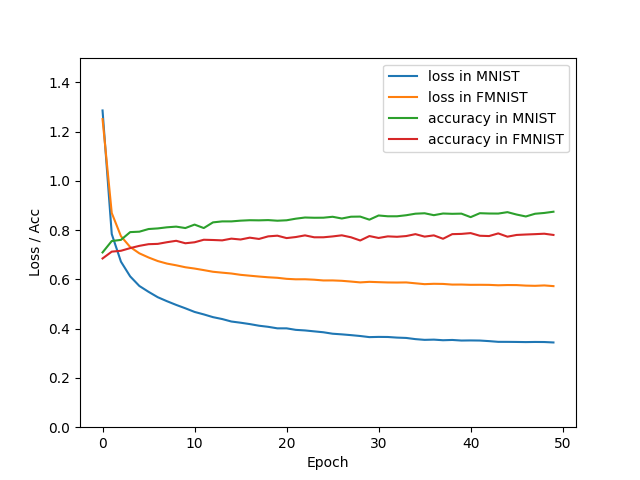}
\caption{Training Performance and Comparison of the New Model.
The figure illustrates the training process and accuracy variations of the new model on the two datasets. It shows the training performance and accuracy changes of the new model on both the MNIST and FashionMNIST datasets.}\label{fig6}
\end{figure}

\begin{figure}[h]%
\centering
\begin{minipage}[c]{0.5\textwidth}
\centering
\includegraphics[height=5cm,width=7cm]{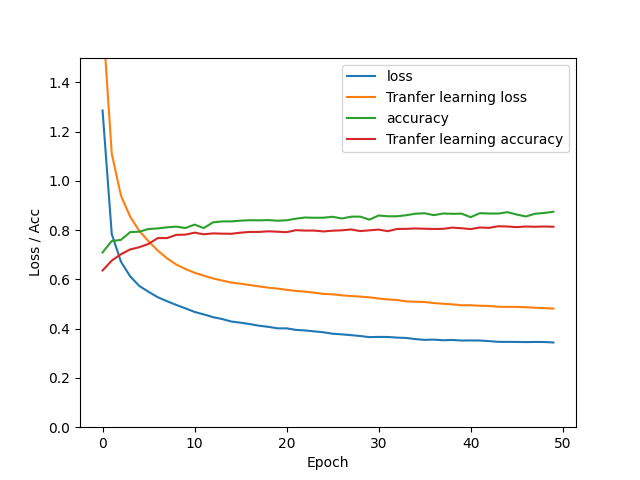}
\end{minipage}%
\begin{minipage}[c]{0.5\textwidth}
\centering
\includegraphics[height=5cm,width=7cm]{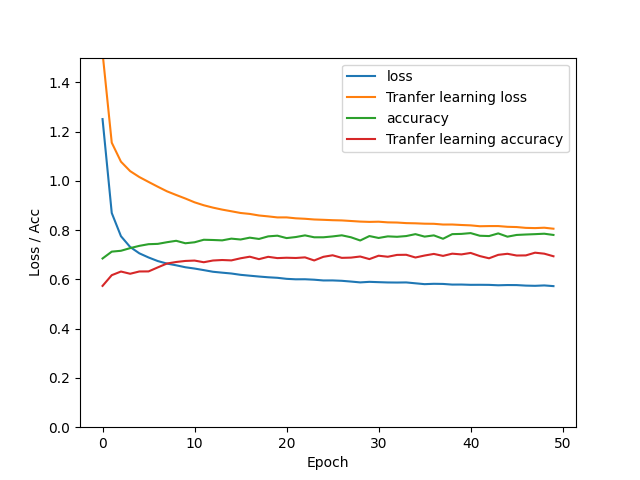}
\end{minipage}
\caption{Comparison between Transfer Learning and Non-Transfer Learning for the New Model.The figure presents a comparison between the training performance of the new model using transfer learning and non-transfer learning approaches on both the MNIST and FashionMNIST datasets. The left graph displays the training results for the MNIST dataset, while the right graph shows the training results for the FashionMNIST dataset.}\label{fig7}
\end{figure}

According to the model described in Section~\ref{subsec31}, we conducted training and testing of PCA-generated low-dimensional data at 8 and 10 dimensions for multi-classification tasks, as shown in Table~\ref{tab1}. In the experiments, we first compared the use of PCA for dimensionality reduction to obtain different low-dimensional datasets to investigate its impact on the QNN classification problem. To demonstrate the model's multi-classification performance, we conducted not only binary classification comparisons but also 8-class classification comparisons. The results indicate that the HQNN model used for binary classification, based on PCA dimensionality reduction, exhibited moderate convergence, with accuracy hovering around 80\% regardless of whether the PCA reduced the dimensions to 8 or 10. However, when faced with the 8-class classification problem, the model encountered a training issue due to the vanishing gradient problem, commonly known as the "barren plateau" problem. The performance on the FashionMNIST dataset mirrored that on the MNIST dataset, as illustrated in Figure ~\ref{fig5}, which solely presents the loss reduction process and accuracy variation for the MNIST dataset. It is evident that the PCA-based HQNN model has demonstrated significant drawbacks in the context of multi-classification problems.

Subsequently, we trained and tested the new model proposed in Section\ref{subsec33} and compared it with the PCA-based approach, as shown in Table~\ref{tab1}. Both models were utilized for the 10-classification task. Leveraging the CNN-based architecture, the new model not only exhibited excellent training performance but also achieved an accuracy of approximately 80\%. In Figure ~\ref{fig6}, we presented the training process and accuracy variation curves for the new QNN model using CNN dimensionality reduction and amplitude encoding on both the MNIST and FashionMNIST datasets. The model achieved an accuracy of around 85\% on the MNIST dataset and about 78.2\% on the FashionMNIST dataset. This demonstrates that the new model, employing CNN-based dimensionality reduction and amplitude encoding, is significantly more effective than the PCA-based approach with angle encoding in multi-classification problems. Furthermore, we compared the new model with the widely used PQC-based QNN model on both datasets. The results show that our proposed model not only outperforms the conventional model during the training process but also achieves an accuracy improvement of approximately 5\%. Therefore, our proposed model proves to be more practical and efficient for image multi-classification tasks.  

In the aforementioned new model, we employed a non-transfer learning approach, while transfer learning might potentially enhance the training process\cite{mari2020transfer}. Transfer learning is an effective method for training new models\cite{bengio2012deep, long2015learning}, allowing the utilization of existing models and knowledge to improve performance in new tasks or domains, while also saving training time and resource costs. In transfer learning, a pre-trained model is typically used as the initial model, which has been trained on a large-scale dataset. Subsequently, the weight parameters of this model can be used as a starting point for fine-tuning or further training on the new task. The benefit of this approach is that the pre-trained model has already learned general feature representations, aiding in faster convergence to the optimal solution for the new task. Hence, we initially trained a standalone CNN model, and then employed transfer learning by fixing the parameters of the convolutional blocks and only training the subsequent QNN layers and fully connected layers, maintaining the same model architecture as the new model. The use of the pre-trained and fixed convolutional blocks can be likened to utilizing an effective feature extractor. A comparison between transfer learning and non-transfer learning for the new model is illustrated in Figure ~\ref{fig7}. Notably, the non-transfer trained model exhibited an approximately 5\% higher accuracy compared to the transfer learning approach. The underlying cause of this phenomenon might be attributed to the fact that the introduced QNN layer is less sensitive to the reduced-dimensional data from the original convolutional blocks when trained in isolation, as opposed to training the entire model together.

\section{Conclusion}\label{sec5}
 In this paper, our main objective was to explore a more practical and efficient quantum neural network (QNN) algorithm by comparing different HQNN model construction approaches in image multi-classification problems. Through our research, we found that the HQNN model using PCA as the dimensionality reduction method encountered the issue of the vanishing gradient problem in multi-classification problems, while the combination of CNN and QNN effectively avoided this problem. Moreover, our newly proposed QNN construction method, which combines amplitude encoding and a more efficient PQC model, demonstrated better performance in image multi-classification problems. However, the transfer learning approach applied to the HQNN model that combines CNN and QNN did not prove to be suitable for this type of problem, or transfer learning might not be appropriate for such HQNN models. Although our model outperformed traditional HQNN models in terms of performance, its accuracy still fell short compared to excellent CNN models in traditional algorithms.

In conclusion, quantum machine learning holds great promise and is expected to bring revolutionary changes to the field of machine learning. By optimizing traditional algorithms and handling feature correlations that classical models cannot perfectly represent, quantum machine learning has the potential to improve the performance of various applications. While there are still technical challenges and not all fields may benefit from quantum computing, with more research and advancements, we believe quantum machine learning will unleash tremendous potential for future scientific and technological development.

\bibliography{sn-bibliography}

\end{document}